\documentclass[a4paper,fleqn]{cas-dc}
\usepackage[numbers,sort&compress]{natbib}
\usepackage{amsmath,amssymb,booktabs,graphicx,tabularx,xcolor}
\usepackage{needspace}
\usepackage{siunitx} 
\newif\ifmarkchanges
\markchangestrue
\definecolor{revisionblue}{RGB}{0,70,150}
\definecolor{revisionred}{RGB}{150,70,0}
\definecolor{revisiongreen}{RGB}{0,120,60}

\newcommand{\revisioncolor}{\ifmarkchanges\color{revisionblue}\fi}

\AddToHook{env/Abstract/begin}{\normalcolor}
\newcommand{\PU}{\operatorname{PU}}
\newcommand{\softplus}{\operatorname{softplus}}
\newcommand{\RMSE}{\operatorname{RMSE}}
\newcommand{\MAE}{\operatorname{MAE}}
\newcommand{\Bias}{\operatorname{Bias}}

\graphicspath{{figures/}}
\begin{document}
\let\WriteBookmarks\relax
\shorttitle{Product-unit GRUs for nuclear masses and charge radii}
\shortauthors{Li et al.}
\title[mode=title]{Product units in gated recurrent networks enhance prediction of nuclear masses and charge radii}
\author[1,2]{Ziyuan Li}
\author[3]{Paulo S. A. Freitas}
\author[3,4]{John W. Clark}
\author[1]{Babette Dellen}
\cormark[1]\ead{dellen@hs-koblenz.de}
\affiliation[1]{organization={Department of Mathematics, Informatics and Technology, University of Applied Sciences Koblenz},addressline={Joseph-Rovan-Allee 2},postcode={53424},city={Remagen},country={Germany}}
\affiliation[2]{organization={Technical University of Munich},city={Munich},country={Germany}}
\affiliation[3]{organization={Department of Mathematics, University of Madeira},addressline={Campus Universit\'{a}rio da Penteada},postcode={9020-105},city={Funchal},country={Portugal}}
\affiliation[4]{organization={Department of Physics, Washington University in St. Louis},addressline={1 Brookings Drive},postcode={63130},city={St. Louis, MO},country={USA}}
\cortext[1]{Corresponding author}

\begin{abstract}
The prediction of nuclear properties using machine learning can complement theoretical models and advance the exploration of poorly known domains of the nuclear chart. We propose a machine learning technique for predicting nuclear masses and charge radii based on gated recurrent units (GRU) that integrates learnable product-unit mappings into the candidate state to model multiplicative dependencies among neighboring nuclei. Computations are performed in the complex domain to jointly capture amplitude and phase dynamics. Across three data partitions and three training seeds per partition, AM-PU-GRU achieves a mean mass interpolation RMSE of 0.2698 MeV and a temporal RMSE of 0.1905 MeV. Configuration comparisons support the integration of learned products into gated recurrence. Without mass inputs, MI-PU-GRU predicts charge radii with an RMSE of 0.019947 fm, 14.40\% below the fixed WS$^*$ prior. Matched radius controls further show how candidate fusion shapes the benefit of learned products. Paired errors, nuclide-resolved maps, and inference measurements characterize predictive gains and computational cost. This article is an extended version of our ICCS 2026 conference paper~\cite{li2026conference}.
\end{abstract}
\begin{keywords}
Product-unit neural networks \sep Gated recurrent units \sep Multiplicative interactions
\sep Nuclear masses \sep Nuclear charge radii \sep Complex-valued neural networks
\end{keywords}
\maketitle
\hypersetup{pdftitle={Product units in gated recurrent networks improve prediction of nuclear masses and charge radii},pdfauthor={Ziyuan Li; Paulo S. A. Freitas; John W. Clark; Babette Dellen},pdfsubject={Product-unit modeling of nuclear masses and charge radii}}

\section{Introduction}\label{sec:introduction}

Accurate prediction of nuclear masses is a fundamental problem in nuclear physics, with important implications for nuclear--structure theory, nucleosynthesis pathways, and applications in nuclear energy and astrophysics. Experimental evaluations such as AME2016 \cite{ame2016i,ame2016ii} and AME2020 \cite{ame2020i,ame2020ii} provide precise measurements for many nuclides, yet large regions of the nuclear chart remain inaccessible. This motivates theoretical and data-driven approaches capable of both interpolation in known regions and extrapolation to unknown nuclides.

Traditional mass models, such as the Weizs\"acker--Skyrme Model (Version 4) (WS4) \cite{ws4}, achieve high accuracy through physics-motivated parametrizations, but their predictive power in unexplored regions is limited \cite{jalili2025recurrent}. Machine learning methods have emerged to complement nuclear-mass prediction, notably recurrent neural network (RNN) \cite{jalili2025recurrent}, gated recurrent unit (GRU) \cite{cho2014gru,jalili2025recurrent}, and mixture density network (MDN) \cite{mumpower2022masses,mumpower2023bayesian,li2024atomic}, achieving strong interpolation accuracy within experimentally known regions, sometimes comparable to physics-based models. These approaches benefit from their ability to learn complex nonlinear relationships directly from data.

Another fundamental nuclear property is the nuclear charge radius, more precisely, the root-mean-square radius of the nuclear charge distribution. 
Charge radii serve to benchmark nuclear models and to validate new
measurements~\cite{mlcharge2025}. 
Theoretical models describe charge radii with rms deviations of about
\SIrange{0.02}{0.05}{\femto\meter}~\cite{mtgp2025}. Also here, machine learning 
offers a data-driven alternative. It learns systematic trends directly
from measured radii and can refine the predictions of physical
models~\cite{utama2016radius}.

Building on GRU architectures for nuclear-mass prediction~\cite{cho2014gru,jalili2025recurrent}, a central architectural question is how explicitly learned products can enrich the GRU candidate state beyond its conventional affine transformation and improve predictions of nuclear properties, both masses and charge radii. Standard gates already multiply states elementwise, but learning exponents in the candidate mapping additionally makes the form of feature interactions trainable. The product-unit (PU) \cite{durbin1989product,leerink1994learning,dellen2019function,dellen2024masses,li2024shape,li2025deep,li2025mri,li2026modeling,li2026pu} approach to machine learning was introduced as an alternative to summation-based formulations, enabling multiplicative interactions that provide compact representations of polynomial and power-law relations. More recently, complex-valued PU extensions \cite{dellen2024masses,li2024shape} have been explored, enabling joint modeling of amplitude and phase dynamics.

Building on these advances, we propose complex-valued GRU extensions for nuclear mass and charge-radius prediction. By integrating multiplicative interactions and product-unit transformations into recurrent frameworks, we obtain two models~\cite{li2026conference}, the multiplicative-interaction product-unit GRU (MI-PU-GRU) and the additive-multiplicative product-unit GRU (AM-PU-GRU).

We next evaluate these architectures on interpolation and temporal extrapolation for nuclear masses and on charge-radius prediction, comparing against real-valued and complex-valued baselines, ablations, and architecture-matched controls. Our contributions are:
a) We embed learnable product-unit mappings in the GRU candidate state, developing MI-PU-GRU and AM-PU-GRU for multiplicative modeling within gated recurrence.
b) We apply a common measured-neighbor and theoretical-prior framework to nuclear masses and charge radii, improving prediction of both observables without mass features for the radius task.
c) We evaluate candidate designs through mass-model comparisons and architecture-matched radius controls, relating predictive gains to candidate fusion, error distributions, and computational cost.

This article extends our ICCS 2026 study~\cite{li2026conference} with charge-radius prediction, additional mass baselines, matched radius controls, and error and computational-cost analyses. Further experiments showed that changing data partitions affects prediction errors and model rankings. Unlike the conference paper, we report the mass-prediction results averaged over three data partitions, with three training seeds per partition, to reduce reliance on a single partition.

\section{Related Work}\label{sec:related}
\subsection{Product units}
Product units (PUs) were introduced as an alternative to summation-based neurons in neural networks. Instead of computing a weighted sum followed by a nonlinear activation, a PU models multiplicative interactions by raising each input to a learnable exponent and taking their product according to
\begin{equation}
y = \prod_{i=1}^n x_i^{w_i}
  = \exp\left(\sum_{i=1}^n w_i\log x_i\right),
\qquad x_i>0.
\label{eq:pu_real}
\end{equation}
This formulation provides a compact representation of polynomial, power-law, and rational relationships. The exponents are learned from data. PU-based architectures have been applied to function and pattern extrapolation~\cite{dellen2019function}, nuclear-mass prediction~\cite{dellen2024masses}, and shape completion~\cite{li2024shape}.

More recently, complex-valued PU networks have been explored for signals with amplitude and phase. In Eq.~\eqref{eq:pu_real}, the complex extension combines $\log|x_i|$ with $\arg(x_i)$ in the exponent, introducing both amplitude scaling and phase rotation via the exponential mapping. Applications include MRI reconstruction and nuclear-mass prediction~\cite{li2025mri,dellen2024masses}.

\subsection{Nuclear models and learned refinements.}
Global mass models supply complementary physical descriptions: the finite-range droplet model combines macroscopic and microscopic terms~\cite{moller2016frdm}, Skyrme--Hartree--Fock--Bogoliubov models use an energy-density-functional framework~\cite{goriely2013hfb}, and WS4 includes a surface-diffuseness correction~\cite{ws4}. Machine learning connects these descriptions with evaluated data~\cite{boehnlein2022review}. Refinements include radial basis functions and Garvey--Kelson relations~\cite{wang2011rbf}, regularized kernel ridge regression~\cite{wu2020krr}, Bayesian neural corrections~\cite{utama2017mass,qu2025mass}, and mirror-symmetry-informed RBF predictions~\cite{li2026mirror}. Direct predictors include probabilistic MDNs~\cite{lovell2022mdn}, physically interpretable mass networks~\cite{mumpower2022masses}, and recurrent models~\cite{jalili2025recurrent}.

\subsection{Charge-radius prediction.}
Evaluated radii~\cite{angeli2013} characterize nuclear size and its isotopic evolution. They are tabulated for roughly 900 isotopes, mostly from laser spectroscopy, muonic spectra, and electron scattering~\cite{angeli2013}. A phenomenological radius formula combines size and isospin terms with WS$^*$ shell corrections and deformations~\cite{wsradius}. Learned approaches include Bayesian refinements of density-functional predictions~\cite{utama2016radius}, feed-forward networks with nuclear-structure features~\cite{wu2020radius}, and pairing-, shell-, isospin-, and shape-informed Bayesian models~\cite{dong2023radius}. Such predictions with uncertainties can guide the choice of new measurements~\cite{giacalone2026}. Kernel models refine phenomenological radii~\cite{ma2022radius} and incorporate odd--even effects~\cite{ma2024radius}. A support vector regression model reports a deviation of \SI{0.016}{\femto\meter} from 10,022 WS$^*$ theoretical radii in an extrapolation comparison~\cite{svr2024}. LightGBM and sparse variational Gaussian processes obtain \SI{0.0121}{\femto\meter} and \SI{0.0142}{\femto\meter} on 24 newly measured nuclei~\cite{lightgbm2025}. Multi-task learning has also been investigated. The NuCLR model reports an rms deviation of \SI{0.0147}{\femto\meter} in out-of-fold validation~\cite{giacalone2026}, building on the original multi-task architecture~\cite{kitouni2023}. A multi-task Gaussian process that learns masses and radii jointly reports a test rms deviation of \SI{0.012}{\femto\meter} with 12 input features~\cite{mtgp2025}. These values rely on different data sets and validation schemes, so they are not directly comparable. Performance also drops for light nuclei and for shape-coexistence regions such as the Hg isotopes~\cite{mlcharge2025}.

\subsection{Multiplicative and complex recurrent representations.}
Learnable arithmetic architectures make algebraic structure explicit: NALU selects arithmetic operations~\cite{trask2018nalu}, NAU/NMU specialize in addition and multiplication~\cite{madsen2020nau}, and Neural Power Units use complex arithmetic to learn powers~\cite{heim2020npu}. Multiplicative integration couples projected recurrent inputs and states~\cite{wu2016mi}; PUs additionally learn exponents. Complex networks couple real and imaginary representations~\cite{trabelsi2018complex}, while unitary recurrences and complex gated networks address state evolution and gradient propagation~\cite{arjovsky2016unitary,wolter2018cgru}. Our method embeds phase-retaining product mappings in the recurrent candidate state to model nuclear neighborhoods.

\section{Methodology}\label{sec:methods}

\subsection{Baseline: Gated Recurrent Unit}
The GRU is a recurrent neural network variant designed to capture long-term dependencies by mitigating the vanishing gradient problem. Two gating mechanisms are introduced: an update gate and a reset gate, which regulate the information flow and memory update within each recurrent unit.

Given an input vector $x_t\in\mathbb R^d$ and the previous hidden state $h_{t-1}\in\mathbb R^H$ at time step $t$, a GRU computes the update gate $z_t\in\mathbb R^H$, reset gate $r_t\in\mathbb R^H$, candidate hidden state $\tilde h_t\in\mathbb R^H$, and the new hidden state $h_t\in\mathbb R^H$ as follows:
\begin{align}
z_t&=\sigma(W_zx_t+U_zh_{t-1}),\label{eq:gru_update}\\
r_t&=\sigma(W_rx_t+U_rh_{t-1}),\\
\tilde h_t&=\tanh(W_hx_t+U_h(r_t\odot h_{t-1})),\label{eq:gru_cand}\\
h_t&=(1-z_t)\odot h_{t-1}+z_t\odot\tilde h_t.\label{eq:gru_hidden}
\end{align}
Here, $\sigma(\cdot)$ denotes the sigmoid activation function, $\tanh(\cdot)$ is the hyperbolic tangent, and $\odot$ represents element-wise multiplication. The learnable parameters satisfy $W_{\{z,r,h\}}\in\mathbb R^{H\times d}$ and $U_{\{z,r,h\}}\in\mathbb R^{H\times H}$. Bias terms are used in our implementation but omitted for notational simplicity.

\begin{figure*}[t]
\centering
\begin{minipage}[t]{0.48\textwidth}
\centering
\textcolor{black}{\includegraphics[width=\linewidth]{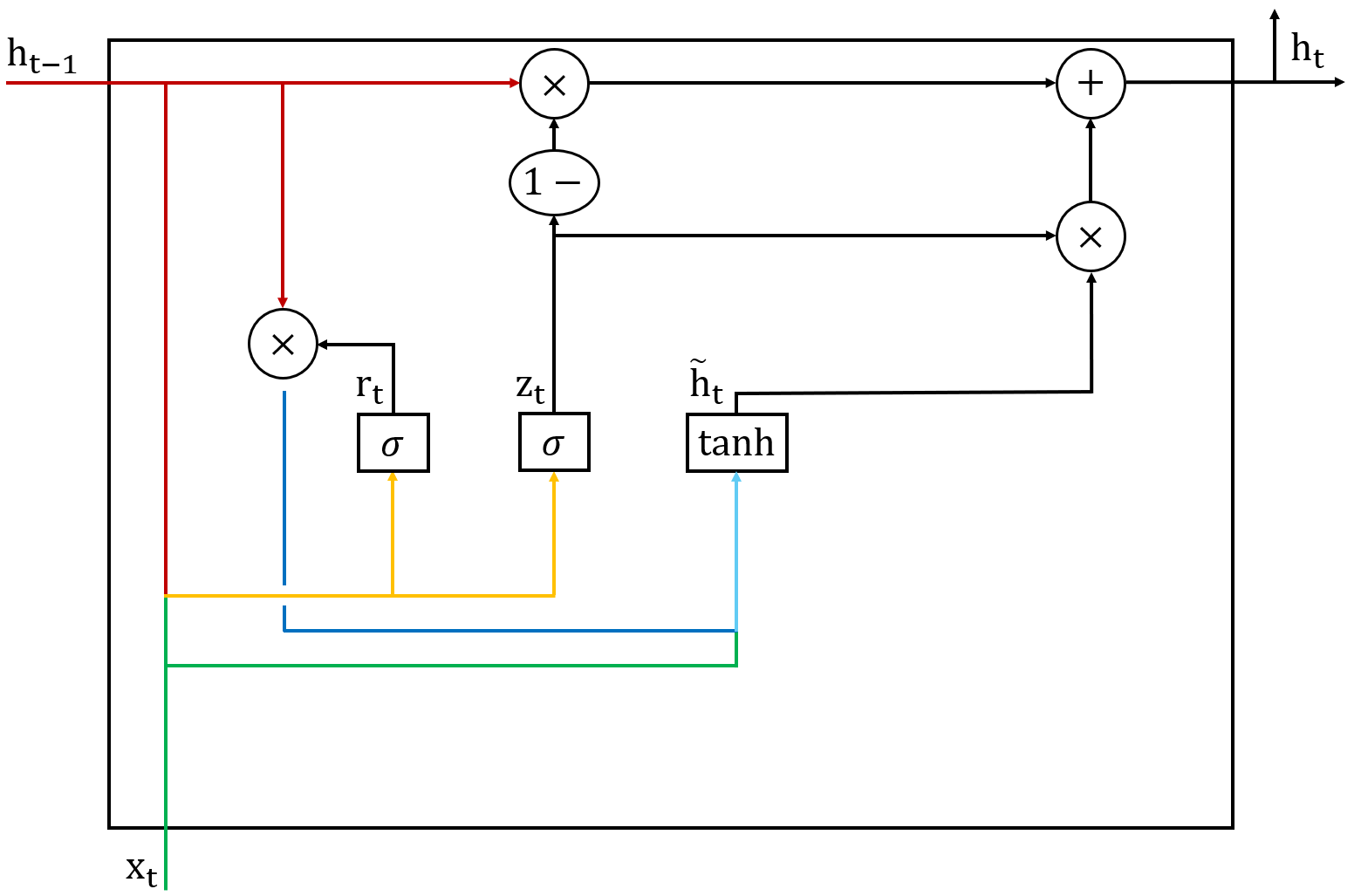}}
\par\smallskip{\fontsize{7.5}{9}\selectfont\mdseries (a) Standard GRU\par}
\end{minipage}\hfill
\begin{minipage}[t]{0.48\textwidth}
\centering
\textcolor{black}{\includegraphics[width=\linewidth]{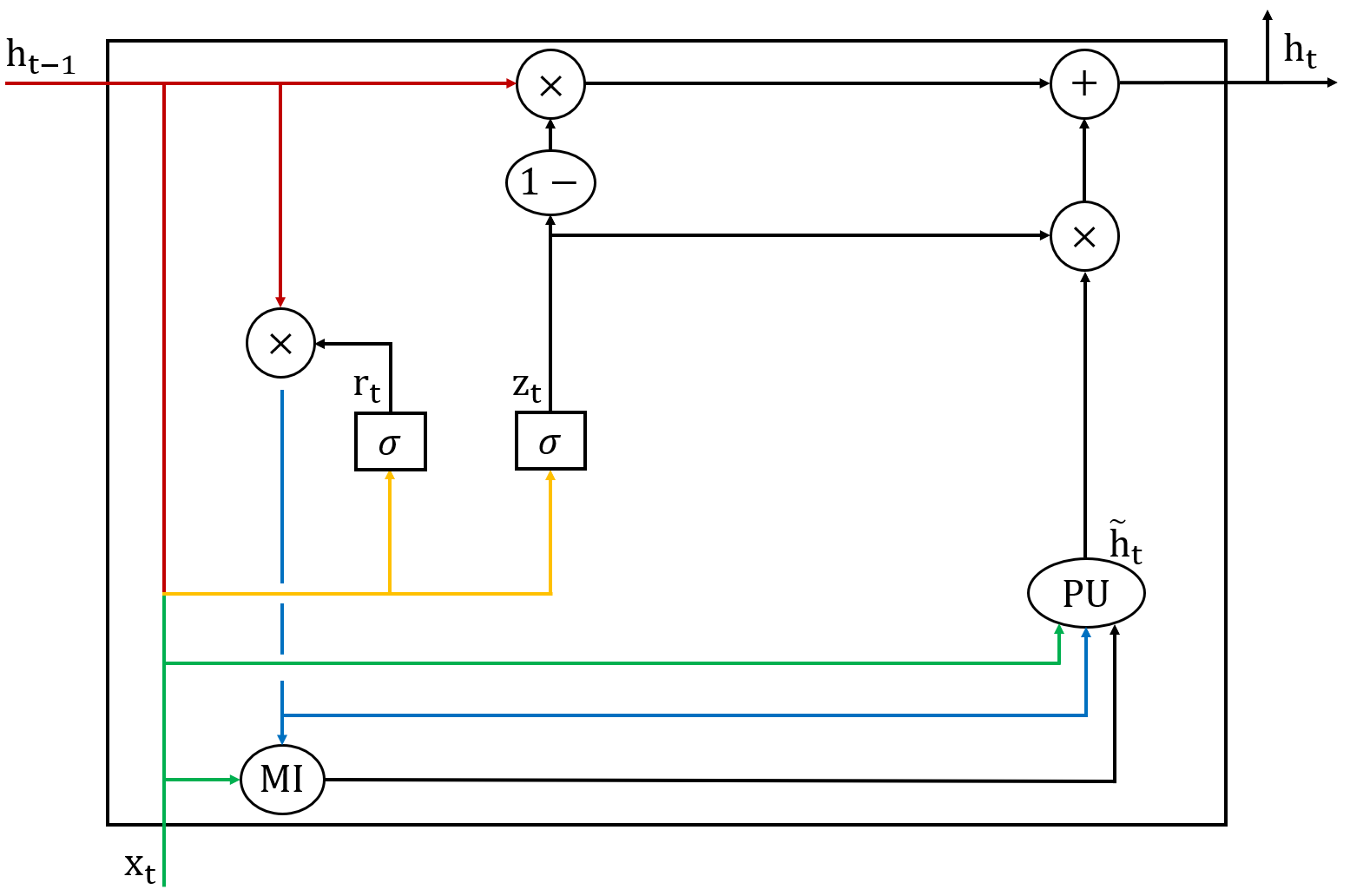}}
\par\smallskip{\fontsize{7.5}{9}\selectfont\mdseries (b) MI-PU-GRU\par}
\end{minipage}
\par\medskip
\begin{minipage}[t]{0.48\textwidth}
\centering
\textcolor{black}{\includegraphics[width=\linewidth]{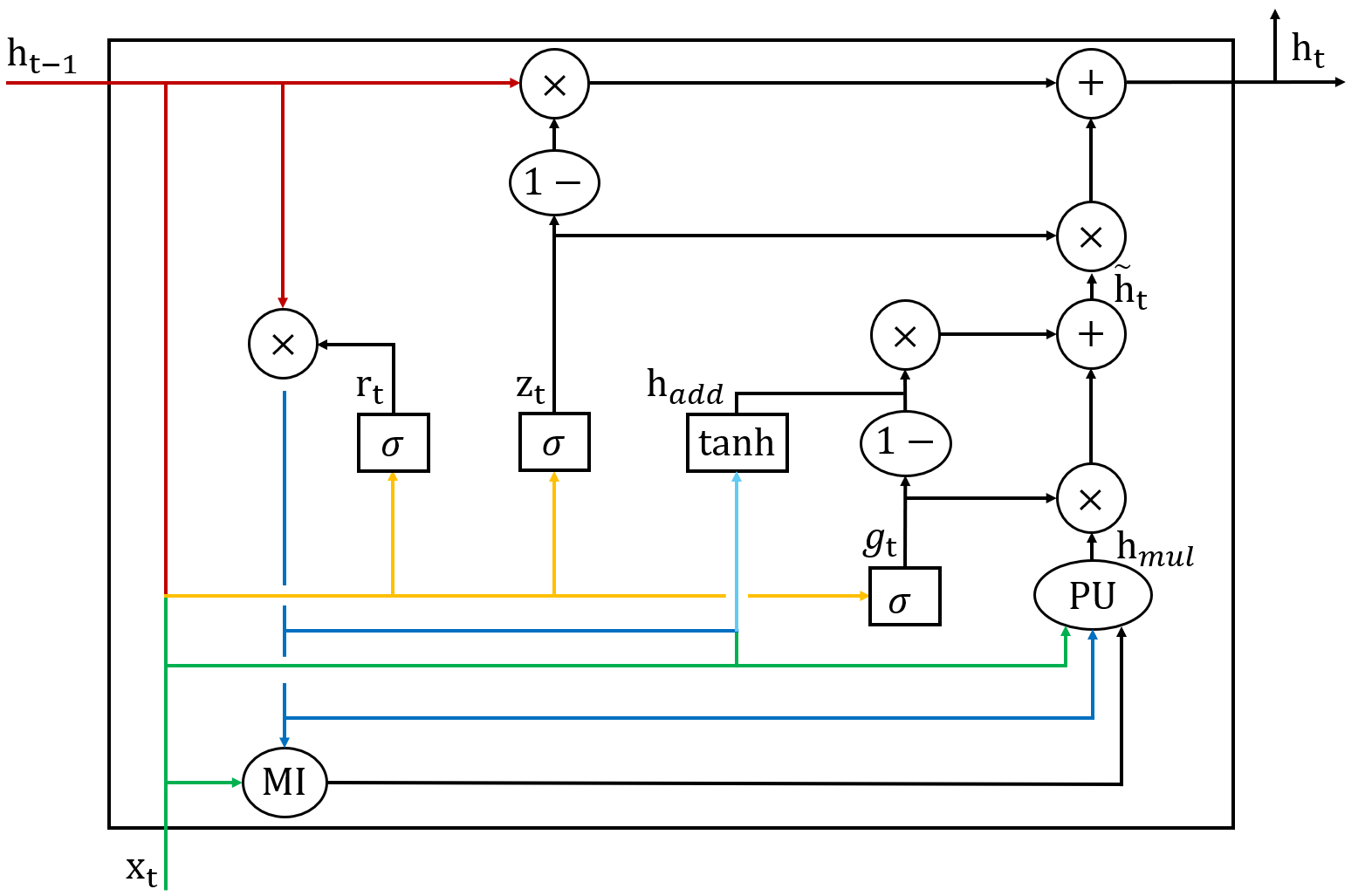}}
\par\smallskip{\fontsize{7.5}{9}\selectfont\mdseries (c) AM-PU-GRU\par}
\end{minipage}
\caption{Computational graphs of (a) a standard GRU cell, (b) MI-PU-GRU, and (c) AM-PU-GRU. Circles marked $\times$ denote element-wise multiplication; projections and biases are implicit. Complex models use real-valued gates and complex recurrent states.}
\label{fig:method}
\end{figure*}

Figure~\ref{fig:method}(a) illustrates the internal structure of a standard GRU cell. The reset gate $r_t$ controls how much of $h_{t-1}$ contributes to the candidate activation $\tilde h_t$. The final hidden state $h_t$ is computed as a componentwise convex combination of $h_{t-1}$ and $\tilde h_t$, governed by the update gate $z_t$. Our PU architectures retain this update mechanism and redesign the candidate computation.

\subsection{Multiplicative-Interaction Product-Unit GRU (MI-PU-GRU)}\label{sec:mi}
To enhance the nonlinear modeling capacity of GRUs, we introduce the MI-PU-GRU cell, which incorporates two components: a multiplicative interaction (MI) branch and a product-unit (PU) transformation in the candidate-state computation. Figure~\ref{fig:method}(b) shows the MI-PU-GRU computational graph.

The MI-PU-GRU case maintains the standard GRU gating structure but modifies the candidate hidden state $\tilde h_t$ as follows:
\begin{align}
z_t&=\sigma(W_zx_t+U_zh_{t-1}),\\
r_t&=\sigma(W_rx_t+U_rh_{t-1}),\\
\mathrm{MI}_t&=\mathrm{MI}(x_t,r_t\odot h_{t-1}),\\
\tilde h_t&=\mathrm{PU}([x_t,r_t\odot h_{t-1},\mathrm{MI}_t]),\\
h_t&=(1-z_t)\odot h_{t-1}+z_t\odot\tilde h_t.
\end{align}
The MI term is computed using an element-wise exponential interaction between the input and hidden representations:
\begin{equation}
\mathrm{MI}(x,h)=\exp\big((W_{\mathrm{mi}}x)\odot(U_{\mathrm{mi}}h)+b_{\mathrm{mi}}\big),
\label{eq:mi}
\end{equation}
where $W_{\mathrm{mi}}$ and $U_{\mathrm{mi}}$ are learnable projection matrices and $b_{\mathrm{mi}}$ is a bias term. The MI branch computes a feature-wise exponential interaction between the input $x_t$ and the reset-gated previous state $r_t\odot h_{t-1}$, producing a multiplicative representation $\mathrm{MI}_t$. The exponential function ensures that the output is strictly positive. This representation, along with $x_t$ and $r_t\odot h_{t-1}$, is fed into a Product-Unit (PU) transformation with concatenated input $v=[x_t,r_t\odot h_{t-1},\mathrm{MI}_t]$:
\begin{align}
\tau&=\softplus(\theta)+10^{-7},\\
\PU_{\mathbb R}(v)
 &=\exp\big(W_{\mathrm{pu}}\operatorname{LN}
       [\log(\max(v,\tau))]+b_{\mathrm{pu}}\big).
\label{eq:realpu}
\end{align}
Here, $\theta$ is a learnable parameter defining the positive floor $\tau$, and $\operatorname{LN}$ denotes layer normalization~\cite{ba2016layernorm}. Both stabilize the logarithmic features.

Compared with standard GRUs, MI-PU-GRU introduces higher-order multiplicative feature interactions and a PU-based nonlinear transformation.

\subsection{Additive-Multiplicative Product-Unit GRU (AM-PU-GRU)}\label{sec:am}
The AM-PU-GRU architecture further extends the MI-PU-GRU strategy by explicitly modeling both additive and multiplicative candidate paths. It introduces a learnable fusion gate to adaptively combine the two contributions. Figure~\ref{fig:method}(c) shows the two-path candidate computation and fusion. The overall update equations are as follows:
\begin{align}
z_t&=\sigma(W_zx_t+U_zh_{t-1}),\\
r_t&=\sigma(W_rx_t+U_rh_{t-1}),\\
g_t&=\sigma(W_gx_t+U_gh_{t-1}),\\
h_{\mathrm{add}}&=\tanh(W_{\mathrm{add}}x_t+U_{\mathrm{add}}(r_t\odot h_{t-1})),\\
\mathrm{MI}_t&=\mathrm{MI}(x_t,r_t\odot h_{t-1}),\\
h_{\mathrm{mul}}&=\mathrm{PU}([x_t,r_t\odot h_{t-1},\mathrm{MI}_t]),\\
\tilde h_t&=g_t\odot h_{\mathrm{mul}}+(1-g_t)\odot h_{\mathrm{add}},\\
h_t&=(1-z_t)\odot h_{t-1}+z_t\odot\tilde h_t.
\label{eq:am_real}
\end{align}
Here, the additive candidate path retains a GRU-style formulation, while the multiplicative candidate is constructed using the MI and PU modules. The fusion gate $g_t$ dynamically balances the two components. The update gate $z_t$ then controls how strongly the fused candidate replaces the previous state.

\subsection{Complex-Valued MI-PU-GRU}\label{sec:pu}
To extend MI-PU-GRU to the complex domain, we adopt a complex-state formulation. The projection weights and recurrent states are complex-valued, while the gates and magnitude thresholds remain real. The real-valued input $x_t\in\mathbb R^d$ is embedded into the complex vector space via a zero-imaginary extension $\tilde x_t=x_t+\mathrm{i}0\in\mathbb C^d$, and the hidden state satisfies $h_{t-1}\in\mathbb C^H$.

The update and reset gates are kept real-valued by applying the sigmoid to the real part of complex pre-activations, yielding $z_t,r_t\in[0,1]^H$:
\begin{align}
z_t&=\sigma\big(\Re(W_z\tilde x_t+U_zh_{t-1})\big),\\
r_t&=\sigma\big(\Re(W_r\tilde x_t+U_rh_{t-1})\big),\\
\mathrm{MI}_t&=\mathrm{MI}_{\mathbb C}(\tilde x_t,r_t\odot h_{t-1}),\\
\tilde h_t&=\mathrm{PU}_{\mathbb C}([\tilde x_t,r_t\odot h_{t-1},\mathrm{MI}_t]),
\label{eq:mi_candidate}\\
h_t&=(1-z_t)\odot h_{t-1}+z_t\odot\tilde h_t.
\label{eq:update}
\end{align}
Using real-valued gates (bounded in $[0,1]$) provides componentwise mixing of complex hidden states without introducing additional phase rotations. Each gate scales the real and imaginary components equally. The MI branch performs feature-wise exponential interactions in complex domain:
\begin{equation}
\mathrm{MI}_{\mathbb C}(x,h)
 =\exp\big((W_{\mathrm{mi}}x)\odot(U_{\mathrm{mi}}h)+b_{\mathrm{mi}}\big),
\end{equation}
where all parameters of this interaction are complex-valued. Since $\exp(a+\mathrm{i}b)=\exp(a)(\cos b+\mathrm{i}\sin b)$, the interaction introduces both amplitude scaling and phase rotation, enhancing expressiveness.

The complex-domain PU transformation is defined as
\begin{equation}
\mathrm{PU}_{\mathbb C}(x)
 =\exp\big(W_{\mathrm{pu}}(\log r(x)+\mathrm{i}\phi(x))+b_{\mathrm{pu}}\big),
\label{eq:complexpu}
\end{equation}
where $r(x)=\max(|x|,\softplus(\theta)+\epsilon)$, $\phi(x)=\arg(x)$, and $\epsilon=10^{-7}$. Here, the logarithm is applied to the stabilized magnitude $r(x)$ and the phase term $\phi(x)$ is handled explicitly. The principal phase retains the sign of negative real inputs, allowing signed standardized features to enter the product representation. The thresholds are real, trainable input-channel parameters.

Since our downstream task requires real-valued predictions, a complex affine readout is followed by a real projection:
\begin{equation}
\widehat y=\Re(w_oh_5+b_o).
\label{eq:readout}
\end{equation}
Both complex PU architectures start from $h_0=\mathbf1+\mathrm{i}\mathbf0$.

\subsection{Complex-Valued AM-PU-GRU}
As with standard AM-PU-GRU, complex AM-PU-GRU builds upon complex MI-PU-GRU by integrating an additive path and a fusion gate to balance additive and multiplicative candidates in $\mathbb C$. The update procedure is given by:
\begin{align}
z_t&=\sigma\big(\Re(W_z\tilde x_t+U_zh_{t-1})\big),\\
r_t&=\sigma\big(\Re(W_r\tilde x_t+U_rh_{t-1})\big),\\
g_t&=\sigma\big(\Re(W_g\tilde x_t+U_gh_{t-1})\big),\\
h_{\mathrm{add}}&=\tanh\big(\Re(W_{\mathrm{add}}\tilde x_t
                +U_{\mathrm{add}}(r_t\odot h_{t-1}))\big),\\
\mathrm{MI}_t&=\mathrm{MI}_{\mathbb C}(\tilde x_t,r_t\odot h_{t-1}),\\
h_{\mathrm{mul}}&=\mathrm{PU}_{\mathbb C}([\tilde x_t,r_t\odot h_{t-1},\mathrm{MI}_t]),\label{eq:am}\\
\tilde h^{\mathrm{add}}_t&=h_{\mathrm{add}}+\mathrm{i}0,\\
\tilde h_t&=g_t\odot h_{\mathrm{mul}}+(1-g_t)\odot\tilde h^{\mathrm{add}}_t,\\
h_t&=(1-z_t)\odot h_{t-1}+z_t\odot\tilde h_t.
\end{align}
Here, $h_{\mathrm{add}}\in\mathbb R^H$ is recast to $\mathbb C^H$ before fusion. The multiplicative path supplies complex candidate information, while the additive candidate and fusion gate remain real-valued.

\subsection{Task Formulation}\label{sec:neighborhood}
Our goal is to predict the mass excess or charge radius of an atomic nucleus from sequential nuclear structure data, which we formulate as a sequence-to-one regression problem. For each target nucleus with proton number $Z^\star$ and neutron number $N^\star$, we construct a short ordered sequence of $T=5$ neighboring nuclides based on proximity in the $(Z,N)$ chart.

Using mass prediction to describe the construction, for a target nucleus $(Z^\star,N^\star)$ with mass number $A^\star=Z^\star+N^\star$, the first $T-1=4$ elements are selected from the reference set of experimentally measured nuclei. Specifically, excluding the target nucleus itself, we consider nuclei satisfying $A\leq A^\star$ and select the four nearest ones in the $(Z,N)$ plane as context nuclei; their measured mass excess values are used as input features. The final element $x_T$ corresponds to the target nucleus; its mass excess value in the input is initialized by the WS4 mass model (i.e., $\mathrm{ME}_T=\mathrm{ME}^{\mathrm{WS4}}(Z^\star,N^\star)$) to provide a physically informed prior.

Given $X$, the model learns a nonlinear mapping $f:\mathbb R^{T\times3}\rightarrow\mathbb R$ to predict the ground-truth mass excess $y=\mathrm{ME}^{\mathrm{AME}}(Z^\star,N^\star)$ of the target nucleus. This design lets the model exploit local continuity over neighboring nuclides while combining empirical measurements and theoretical priors. For charge radii, the layout uses measured neighbor radii and a WS$^*$ target estimate, without mass features.

In addition to these primary models, we also implement their real-valued counterparts (MI-PU-GRU and AM-PU-GRU) to compare real and complex formulations. Furthermore, we design ablation variants in the complex domain, including PU-GRU, MI-GRU, AM-GRU, and the baseline GRU, by selectively disabling architectural components.

\subsection{Reference Architectures and Matched Controls}\label{sec:controls}
CV denotes complex-valued. Throughout this article, CV GRU denotes the baseline that applies Tanh separately to real and imaginary candidate components:
\begin{equation}
\begin{aligned}
b_t&=W_h\tilde x_t+U_h(r_t\odot h_{t-1})+b_h,\\
\tilde h_t&=\tanh(\Re b_t)+\mathrm{i}\tanh(\Im b_t).
\end{aligned}
\label{eq:cvgru}
\end{equation}
Its real gates, initial state, and complex readout follow the conventions above. Real GRU uses an initial hidden state of ones. CV PU-GRU omits the MI branch and supplies $[\tilde x_t;r_t\odot h_{t-1}]$ to the PU. The CV MI-GRU and AM-GRU configurations replace the multiplicative candidate by Tanh of the real part of a projected interaction; they serve as complete-configuration controls.

Candidate-replacement models substitute the PU in Eqs.~\eqref{eq:mi_candidate} and~\eqref{eq:am} with
\begin{align}
\mathcal A_\phi(v)&=\phi(\Re(W_pv+b_p))
                 +\mathrm{i}\phi(\Im(W_pv+b_p)),\\
\phi(s)&\in\{\tanh(s),\,s\sigma(s)\}.
\label{eq:matched}
\end{align}
The second activation is SiLU~\cite{elfwing2018silu}. Candidate inputs, hidden width, gates, MI interaction, AM fusion, initial state, and readout are preserved. For radii, shared-tensor initialization and optimization are also held fixed, giving architecture-matched PU controls. The mass study compares these candidate models with the full MI-PU-GRU and AM-PU-GRU designs. A same-input MLP receives the standardized sequence flattened to 15 inputs, with two Tanh hidden layers and a scalar linear output.

\section{Data and Evaluation Protocol}\label{sec:protocol}

Both tasks use the neighborhood representation in Section~\ref{sec:neighborhood}, with the data, priors, and evaluation settings below.

\subsection{Nuclear Data and Fixed Priors}\label{sec:data}

\paragraph{Mass data and fixed priors.}
We use experimental mass excesses from the AME2020 evaluation~\cite{ame2020i,ame2020ii}, selecting nuclei with $Z,N\geq8$ and excluding estimated values. The prediction target is mass excess, with errors reported in MeV. The original WS4 mass-excess table~\cite{ws4} provides the fixed theoretical prior.

\paragraph{Charge-radius data and fixed prior.}
We use experimental charge radii from the 2013 evaluation~\cite{angeli2013}, retaining finite positive radii with valid coordinates and $Z,N\geq8$. Radii and their reported measurement uncertainties are in fm; uncertainties are used only for diagnostics. The fixed prior is the original four-parameter charge-radius formula based on WS$^*$ shell corrections and deformations~\cite{wsradius}, without RBF or kernel corrections.

\subsection{Data Splits and Neighborhood Construction}\label{sec:splits}

Models compared within each run ensemble share the same data partitions.

\paragraph{Experiment I: Interpolation.}
This experiment assesses the model's interpolation capability within the domain of experimentally known nuclides. We randomly split the AME2020 dataset into training, validation, and test subsets using a 70\%-15\%-15\% ratio. During sequence construction for the validation and test sets, only the experimentally measured mass-excess values from the training set are used to provide input features for precursor nuclei.

\paragraph{Experiment II: Temporal Extrapolation.}
To test the extrapolation capability of our models, we simulate a temporal generalization scenario by training on nuclei discovered by 2016 and testing on later-discovery nuclei. Both groups use AME2020 mass labels, with discovery-year groups consistent with NUBASE2020~\cite{nubase2020}. The reference pool is split 95/5 into training and interpolation-test sets; later-discovery nuclei form the temporal test. As in Experiment I, all sequences for both extrapolation and interpolation evaluation are constructed using only nuclei from the training set to provide precursor mass excess values.

To assess robustness to data partitioning, we evaluate the mass models in Tables~\ref{tab:mass} and~\ref{tab:mass_checkpoint} with data split seeds 1, 2, and 3 and training seeds 3, 4, and 5 for each split, giving nine runs per model and task. Training seeds vary initialization and training randomness while keeping each partition fixed.

\paragraph{Additional mass comparisons.}
The additional models in Table~\ref{tab:mass_checkpoint} use the same data partitions, measured-neighbor sequences, WS4 input arrays, targets, scalers, and training seeds as the main models in Table~\ref{tab:mass}. Both Tanh and SiLU candidate replacements are evaluated in Mass I and Mass II. The CV GRU and full MI-PU-GRU and AM-PU-GRU results in Table~\ref{tab:mass_checkpoint} are the same nine-run summaries reported in Table~\ref{tab:mass}.

\paragraph{Charge-radius interpolation.}
The radius experiments use a 70\%-15\%-15\% training--validation--test partition. A held-out target is predicted from four measured training-neighbor radii, its coordinates, and its theoretical radius. Inputs contain no measured or predicted mass.

The radius dataset uses split seed 0: the combined held-out fraction is selected first, then divided equally into validation and test subsets using the same seed.

For the radius experiments, measured contexts come exclusively from the initial training partition, using Euclidean distance and the mass-number constraint in Section~\ref{sec:neighborhood}. The pool is sorted by $(Z,N)$, and NumPy's quicksort orders distances, including ties. Targets with fewer than four eligible neighbors are omitted, while the full initial training partition remains the context pool.

Radius inputs are standardized positionwise: the $5\times3$ array is flattened, its 15 means and scales are fitted on effective training sequences, and the array is reshaped afterward. The scalar target is standardized separately. For the mass experiments in Tables~\ref{tab:mass} and~\ref{tab:mass_checkpoint}, scalers are fitted to the training sequences separately for each data split and shared across models and training seeds. Models predict the observable directly, with the theoretical estimate supplied as an input feature.

\subsection{Training and Model Selection}\label{sec:training}
All neural-network models are trained in PyTorch for 3000 epochs using MSE loss and the RAdam optimizer \cite{liu2019radam} (initial learning rate $10^{-3}$). We use a step learning-rate schedule that halves the learning rate every 500 epochs, and set the batch size to 64.

Each recurrent network applies one cell at five sequence positions. Mass hidden widths are 120 for real GRU, 90/75 for real MI/AM-PU-GRU, 80 for CV GRU and PU-GRU, 60/50 for CV MI/AM-PU-GRU, and 70/60 for CV MI/AM-GRU. Radius widths are half these values, rounded upward. The additional MLP uses $15\to187\to187\to1$ units for mass and $15\to92\to92\to1$ for radii, approximately matching CV MI's parameter count. For complex-valued architectures, each complex parameter is represented as $a+ib$, where $a$ is the real and $b$ the imaginary part, and counts as two real components. Real-valued parameters, including PU thresholds, count once.

Gradient clipping~\cite{pascanu2013gradients} uses a norm threshold of 1; mass models use zero weight decay. Radius PU floors start near 1, with raw thresholds frozen for five epochs before learning. Radius models use RAdam's coupled $L_2$ penalty of $10^{-4}$ on matrix weights, excluding biases, thresholds, and nonmatrix parameters.

Mass I and charge-radius models use the minimum-validation-loss checkpoint. All Mass II models are evaluated at the final training epoch. Additional mass and radius experiments use Tesla V100-SXM2-32GB GPUs, PyTorch 2.7.1, and CUDA 11.8.

\subsection{Evaluation Metrics and Paired Analysis}\label{sec:metrics}
For errors $e_i=\widehat q_i-q_i$, the metrics are
\begin{align}
\RMSE&=\sqrt{n^{-1}\sum_i e_i^2},\\
\MAE&=n^{-1}\sum_i|e_i|,\\
\Bias&=n^{-1}\sum_i e_i.
\end{align}
Metrics are computed per fitted model; tables give run counts and report means with sample standard deviations where available. Radius models use training seeds 0--4. Percentage reductions use unrounded means.

Tables~\ref{tab:mass} and~\ref{tab:mass_checkpoint} report the mean and sample standard deviation of the nine run RMSEs, incorporating both partition and training variability.

Paired radius gains are comparator minus candidate errors. Pointwise 95\% intervals use 50,000 percentile-bootstrap resamples of isotope chains grouped by $Z$, retaining all five seed predictions together~\cite{field2007bootstrap}. The intervals are conditional on the fitted models and partition; shared neighbors may induce cross-chain dependence. Leave-one-nucleus and leave-one-chain analyses assess influence.

\section{Nuclear-Mass Prediction}\label{sec:mass}

We examine mass interpolation and temporal prediction under the WS4 prior, comparing real and complex formulations, GRU baselines, and configuration ablations across three data partitions with three training seeds per partition.

\subsection{Main Results}

To evaluate the predictive capability of the proposed architectures, we first compare the baseline real-valued GRU with the complex-valued MI-PU-GRU and AM-PU-GRU under the WS4 prior.

The interpolation results of Experiment I are summarized in Table~\ref{tab:mass}. Both complex-valued PU-GRU variants have lower mean errors than the real-valued GRU baseline. In particular, the complex AM-PU-GRU model achieves the lowest interpolation error (0.2698 MeV) and the lowest sample standard deviation among these three models (0.0129 MeV), highlighting the effectiveness of combining additive and multiplicative candidate paths within the complex domain.

\begin{table*}[t]
\caption{Nuclear-mass RMSE (MeV): means $\pm$ sample standard deviations over nine runs per model and task, crossing data split seeds 1, 2, and 3 with training seeds 3, 4, and 5. Both Mass II columns report test errors from the final training epoch. Bold marks column minima.}
\label{tab:mass}
\centering\small
\setlength{\tabcolsep}{7pt}
\begin{tabular}{@{}lrrr@{}}
\toprule
Model & Mass I: interpolation & Mass II: interpolation & Mass II: temporal \\
\midrule
\multicolumn{4}{@{}l}{\textit{Main comparisons}}\\
CV MI-PU-GRU & $0.2708 \pm 0.0149$ & $0.2536 \pm 0.0110$ & $0.2251 \pm 0.0389$ \\
CV AM-PU-GRU & $\mathbf{0.2698 \pm 0.0129}$ & $\mathbf{0.2509 \pm 0.0085}$ & $\mathbf{0.1905 \pm 0.0333}$ \\
Real GRU & $0.2943 \pm 0.0179$ & $0.2552 \pm 0.0156$ & $0.2764 \pm 0.0449$ \\
\midrule
\multicolumn{4}{@{}l}{\textit{Complex GRU baseline}}\\
CV GRU & $0.2909 \pm 0.0070$ & $0.2685 \pm 0.0115$ & $0.2673 \pm 0.0233$ \\
\midrule
\multicolumn{4}{@{}l}{\textit{Real-valued counterparts}}\\
Real MI-PU-GRU & $0.3067 \pm 0.0145$ & $0.2737 \pm 0.0119$ & $0.4508 \pm 0.1340$ \\
Real AM-PU-GRU & $0.2885 \pm 0.0081$ & $0.2668 \pm 0.0192$ & $0.4529 \pm 0.1424$ \\
\midrule
\multicolumn{4}{@{}l}{\textit{Configuration ablations}}\\
CV PU-GRU & $0.7820 \pm 0.1245$ & $0.5628 \pm 0.0971$ & $0.4497 \pm 0.3186$ \\
CV MI-GRU & $0.3655 \pm 0.0374$ & $0.2925 \pm 0.0159$ & $0.2716 \pm 0.0448$ \\
CV AM-GRU & $0.3608 \pm 0.0423$ & $0.3018 \pm 0.0105$ & $0.3419 \pm 0.0354$ \\
\bottomrule
\end{tabular}
\end{table*}

The results of Experiment II are summarized in Table~\ref{tab:mass}. Compared to the real-valued GRU baseline, both complex PU-GRU variants yield lower extrapolation errors, with the complex AM-PU-GRU again achieving the best overall performance among the main models. In particular, AM-PU-GRU attains the lowest interpolation RMSE of $0.2509\pm0.0085$ MeV and the lowest extrapolation RMSE of $0.1905\pm0.0333$ MeV. Moreover, AM-PU-GRU exhibits the lowest temporal standard deviation across partitions and training runs among these three models.

\subsection{Comparative Analysis: Real vs. Complex}
To assess the effect of complex-valued modeling, we compare the performance of real-valued PU-GRU variants with their complex-valued counterparts under the WS4 prior. Their hidden widths are chosen to give comparable parameter counts (Section~\ref{sec:training}). Each real-valued network uses the same three data partitions and three training seeds per partition as its complex counterpart; the reported means and sample standard deviations aggregate all nine runs.

The results, summarized in Table~\ref{tab:mass}, indicate that the complex-valued PU-GRU variants have lower mean errors than their real-valued counterparts in both interpolation and extrapolation settings. The complex product pathway retains phase alongside stabilized magnitude, allowing signed input and hidden features to enter the learned representation.

\subsection{Mass-Model Configuration Analysis}\label{sec:mass_config}
To examine the candidate pathways, we compare complete configurations in the complex domain. We implement four variants: a complex-valued GRU, a complex MI-GRU (without PU and additive path), a complex AM-GRU (without PU but with additive--multiplicative fusion), and a complex PU-GRU (retaining PU but without explicit MI or AM mechanisms). Their candidate definitions and hidden widths are given in Sections~\ref{sec:controls} and~\ref{sec:training}.

In Mass I (Table~\ref{tab:mass}), CV GRU, PU-GRU, MI-GRU, and AM-GRU yield mean RMSEs of 0.2909, 0.7820, 0.3655, and 0.3608 MeV, respectively. The full MI-PU-GRU and AM-PU-GRU designs reach 0.2708 and 0.2698 MeV. Thus, the PU-only configuration does not improve on CV GRU, while the complete MI--PU and AM--PU candidate designs have lower mean errors.

For Mass II interpolation, CV GRU, PU-GRU, MI-GRU, and AM-GRU yield test RMSEs of 0.2685, 0.5628, 0.2925, and 0.3018 MeV, respectively. Their temporal errors are 0.2673, 0.4497, 0.2716, and 0.3419 MeV. MI-PU-GRU and AM-PU-GRU improve the interpolation means to 0.2536 and 0.2509 MeV and the temporal means to 0.2251 and 0.1905 MeV. These configuration comparisons favor combining PU with the interaction and fusion pathways.

\section{Nuclear Charge-Radius Prediction}\label{sec:radius}

Charge-radius prediction retains the neighborhood layout and candidate architectures, replacing mass inputs with measured neighbor radii and a theoretical radius prior. We evaluate overall accuracy and the distribution of gains across the nuclear chart.

\begin{table*}[t]

\caption{ Charge-radius test errors (fm). Neural-model results are five-seed means $\pm$ sample standard deviations; the WS$^*$ prior is deterministic. Bold marks the lowest mean RMSE and MAE. Bias is prediction minus measurement.}
\label{tab:radius}
\centering\small
\setlength{\tabcolsep}{8pt}
\begin{tabular}{@{}lrrr@{}}
\toprule
Model & RMSE & MAE & Bias \\
\midrule
\multicolumn{4}{@{}l}{\textit{Main comparisons}} \\
CV MI-PU-GRU & $0.019947 \pm 0.000121$ & $\mathbf{0.012640 \pm 0.000168}$ & $-0.000730 \pm 0.000377$ \\
CV AM-PU-GRU & $0.020758 \pm 0.000517$ & $0.014053 \pm 0.000445$ & $-0.002325 \pm 0.000967$ \\
Real GRU & $0.020399 \pm 0.000163$ & $0.012995 \pm 0.000255$ & $-0.001233 \pm 0.000643$ \\
CV GRU & $0.020740 \pm 0.000068$ & $0.013379 \pm 0.000082$ & $-0.002008 \pm 0.000444$ \\
MLP (same input) & $0.021314 \pm 0.000107$ & $0.013606 \pm 0.000133$ & $-0.000661 \pm 0.000535$ \\
WS$^*$ prior & $0.023303$ & $0.015523$ & $+0.000585$ \\
\midrule
\multicolumn{4}{@{}l}{\textit{Real-valued counterparts}} \\
Real MI-PU-GRU & $0.021385 \pm 0.000891$ & $0.014488 \pm 0.000620$ & $-0.001620 \pm 0.001176$ \\
Real AM-PU-GRU & $0.020372 \pm 0.000865$ & $0.013487 \pm 0.000653$ & $-0.000404 \pm 0.001983$ \\
\midrule
\multicolumn{4}{@{}l}{\textit{Model configuration controls}} \\
CV PU-GRU & $0.023088 \pm 0.000267$ & $0.015929 \pm 0.000313$ & $-0.002731 \pm 0.000672$ \\
CV MI-GRU & $0.024489 \pm 0.000523$ & $0.016785 \pm 0.000280$ & $-0.003203 \pm 0.000584$ \\
CV AM-GRU & $0.025885 \pm 0.001430$ & $0.016947 \pm 0.000625$ & $-0.002758 \pm 0.000934$ \\
\midrule
\multicolumn{4}{@{}l}{\textit{Architecture-matched PU controls}} \\
CV MI (matched Tanh) & $\mathbf{0.019905 \pm 0.000115}$ & $0.012766 \pm 0.000134$ & $-0.001564 \pm 0.000547$ \\
CV AM (matched Tanh) & $0.023060 \pm 0.001017$ & $0.015586 \pm 0.000672$ & $-0.003176 \pm 0.000541$ \\
CV MI (matched SiLU) & $0.020206 \pm 0.000086$ & $0.013017 \pm 0.000098$ & $-0.001326 \pm 0.000390$ \\
CV AM (matched SiLU) & $0.022337 \pm 0.001027$ & $0.014972 \pm 0.000757$ & $-0.001620 \pm 0.000854$ \\
\bottomrule
\end{tabular}
\end{table*}

\subsection{Predictive Accuracy}
On the charge-radius test set (Table~\ref{tab:radius}), CV MI-PU-GRU achieves an RMSE of $0.019947\pm0.000121$ fm and an MAE of $0.012640\pm0.000168$ fm, reducing the WS$^*$ prior errors by 14.40\% and 18.58\%. Mean RMSE is also lower than real GRU, CV GRU, and the same-input MLP by 2.21\%, 3.82\%, and 6.41\%, respectively.

MI's mean bias of $-0.000730$ fm indicates slight overall underprediction. Its lower RMSE and MAE relative to both GRU baselines show gains in squared-error and absolute-error performance.

CV AM-PU-GRU reaches $0.020758\pm0.000517$ fm RMSE, reducing the prior error by 10.92\% and outperforming its matched Tanh and SiLU replacements. MI-Tanh gives a 0.21\% lower RMSE than MI-PU ($0.019905\pm0.000115$ fm), while MI-PU has the lowest mean MAE across all evaluated models. Both PU designs improve on the fixed prior.

\subsection{Nuclide-Resolved Error Analysis}\label{sec:radius_map}

\begin{figure*}[tp]
\centering
\includegraphics[width=0.95\textwidth]{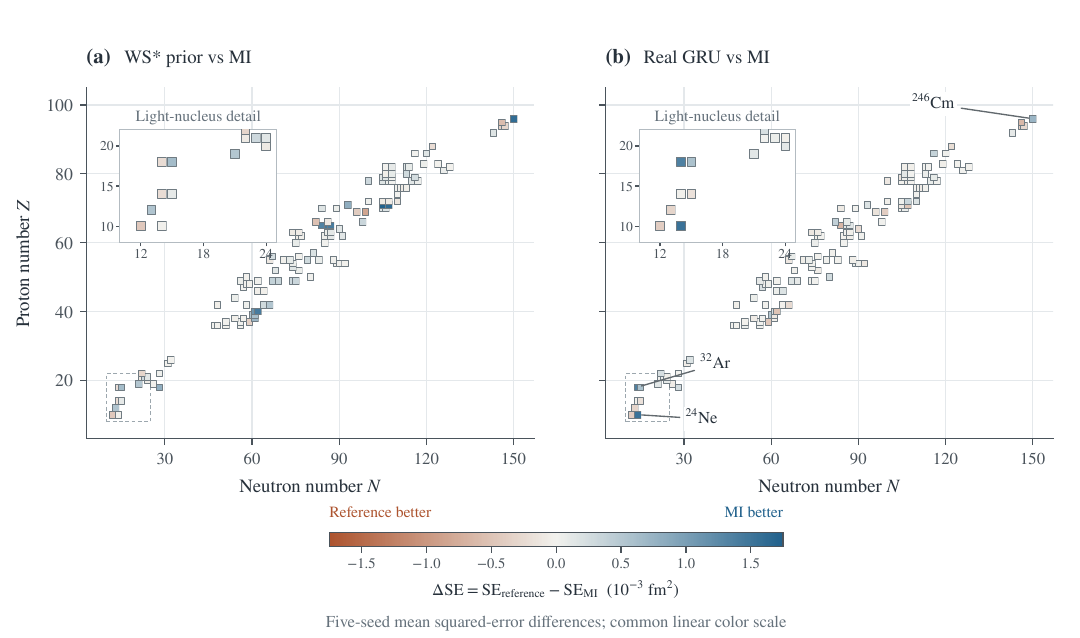}
\caption{ Nuclide-resolved radius gains of CV MI-PU-GRU. Squares mark test nuclei at $(N,Z)$. Color shows comparator minus MI squared error, averaged across seeds: blue favors MI; orange favors the comparator. Left: WS$^*$ prior; right: real GRU.}
\label{fig:nuclides}
\end{figure*}

Figure~\ref{fig:nuclides} locates the radius-error gains across the nuclear chart. MI reduces seed-averaged squared error relative to WS$^*$ for 63.4\% of test nuclei and lowers RMSE in all four mass-number groups: $A<60$, $60\leq A<120$, $120\leq A<180$, and $A\geq180$. Against real GRU, it improves 50.4\% of test nuclei, with gains concentrated more strongly among light nuclei.

MI retains lower mean RMSE than real GRU after omitting any single nucleus or isotope chain. Joint omission of the three largest favorable contributors, $^{24}$Ne, $^{32}$Ar, and $^{246}$Cm, reverses the ordering. Separately, $^{152}$Tb, $^{151}$Tb, and $^{149}$Tb account for 44.6\% of MI's total squared error. The gains thus withstand individual omissions but remain sensitive to a small group of influential nuclei.

\section{Component and Reliability Analysis}\label{sec:analysis}

\begin{figure*}[t]
\centering
\includegraphics[width=0.95\textwidth]{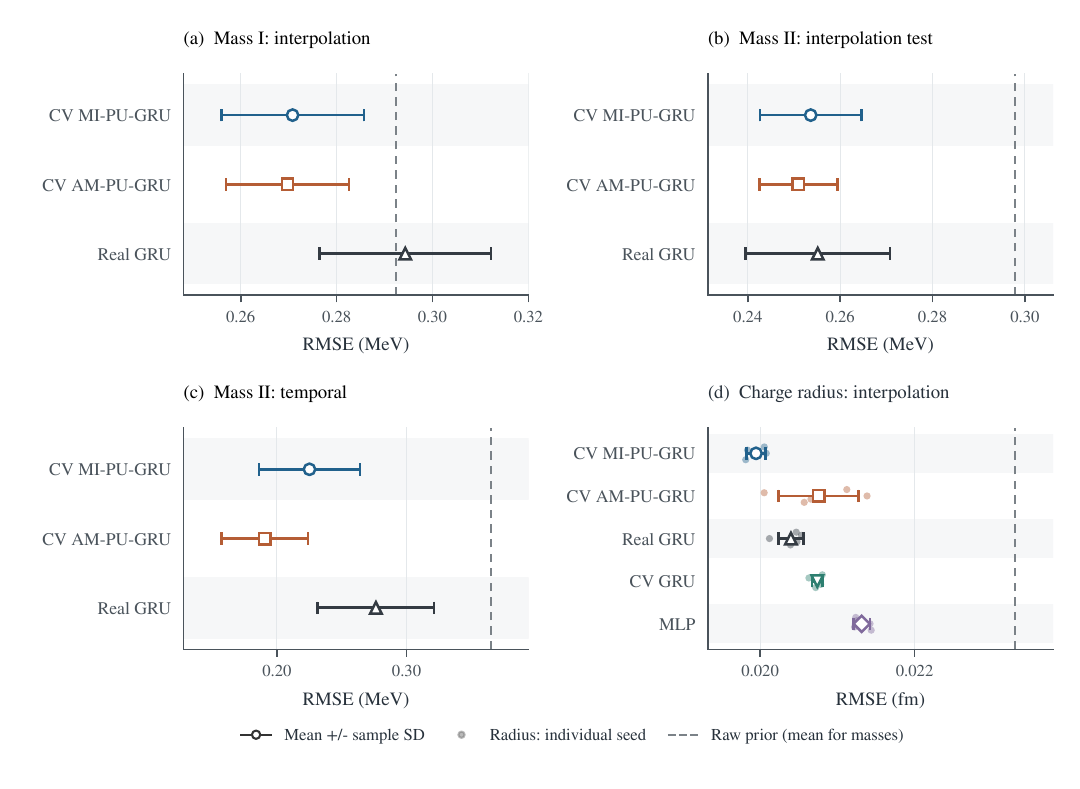}
\caption{Mass and radius prediction errors. Symbols and whiskers show means and sample standard deviations. Mass panels use the nine-run aggregation in Table~\ref{tab:mass}; radius results use five training seeds, shown as small points. Dashed lines show the mean raw WS4 error across the three mass partitions or the fixed radius-prior error. Panel (b) reports Mass II interpolation test errors.}
\label{fig:overview}
\end{figure*}

Figure~\ref{fig:overview} summarizes predictive performance. We next compare additional mass models, isolate the radius PU contribution, and examine paired errors and inference cost.

\subsection{Mass-Model Performance Comparisons}\label{sec:mass_comparison}

Table~\ref{tab:mass_checkpoint} compares the full MI-PU-GRU and AM-PU-GRU designs with CV GRU, a same-input MLP, and Tanh/SiLU candidate replacements using the common inputs and nine-run seed ensemble specified in Section~\ref{sec:splits}. The full PU models and CV GRU are repeated from Table~\ref{tab:mass} to facilitate direct comparison.

\begin{table*}[t]
\caption{Mass-model RMSE (MeV): means $\pm$ sample standard deviations over nine runs per model and task, crossing data split seeds 1, 2, and 3 with training seeds 3, 4, and 5. All models share the input data for each split. Full PU-model and CV GRU results are repeated from Table~\ref{tab:mass}. Both Mass II columns report test errors from the final training epoch. Bold marks column minima.}
\label{tab:mass_checkpoint}
\centering\small
\setlength{\tabcolsep}{7pt}
\begin{tabular}{@{}lrrr@{}}
\toprule
Model & Mass I: interpolation & Mass II: interpolation & Mass II: temporal \\
\midrule
\multicolumn{4}{@{}l}{\textit{Full PU models}}\\
CV MI-PU-GRU & $0.2708 \pm 0.0149$ & $0.2536 \pm 0.0110$ & $0.2251 \pm 0.0389$ \\
CV AM-PU-GRU & $\mathbf{0.2698 \pm 0.0129}$ & $\mathbf{0.2509 \pm 0.0085}$ & $\mathbf{0.1905 \pm 0.0333}$ \\
\midrule
\multicolumn{4}{@{}l}{\textit{Additional baselines}}\\
CV GRU & $0.2909 \pm 0.0070$ & $0.2685 \pm 0.0115$ & $0.2673 \pm 0.0233$ \\
MLP (same input) & $0.3249 \pm 0.0439$ & $0.2841 \pm 0.0110$ & $0.2823 \pm 0.0720$ \\
\midrule
\multicolumn{4}{@{}l}{\textit{Candidate replacements}}\\
CV MI (Tanh) & $0.3553 \pm 0.0525$ & $0.2626 \pm 0.0109$ & $0.2816 \pm 0.0935$ \\
CV AM (Tanh) & $0.3562 \pm 0.0640$ & $0.2582 \pm 0.0061$ & $0.2595 \pm 0.0831$ \\
CV MI (SiLU) & $0.2800 \pm 0.0106$ & $0.2699 \pm 0.0108$ & $0.2270 \pm 0.0286$ \\
CV AM (SiLU) & $0.2837 \pm 0.0049$ & $0.2686 \pm 0.0129$ & $0.2314 \pm 0.0497$ \\
\bottomrule
\end{tabular}
\end{table*}

Both full PU designs have lower mean RMSEs than the additional baselines and candidate replacements in all three evaluations. AM-PU-GRU achieves the lowest means: 0.2698 MeV for Mass I interpolation, 0.2509 MeV for Mass II interpolation, and 0.1905 MeV for temporal prediction. In Mass I, the SiLU replacements are closer to the full PU designs than the Tanh replacements, with mean errors of 0.2800 and 0.2837 MeV for MI and AM, respectively.

For temporal prediction, MI-PU-GRU and MI-SiLU are close, at 0.2251 and 0.2270 MeV, whereas AM-PU-GRU has a lower mean error than both AM-Tanh (0.2595 MeV) and AM-SiLU (0.2314 MeV). The performance differences between the full PU models and their candidate replacements therefore depend on the candidate pathway and evaluation task.

\subsection{Architecture-Matched Radius PU Controls}\label{sec:components}

\begin{table*}[t]
\caption{Architecture-matched radius controls. Replacement RMSEs (fm) are five-seed means $\pm$ sample standard deviations. Reductions and gains compare replacement with PU; positive values favor PU. Pairs count improved seeds. Gain intervals are pointwise 95\% isotope-chain bootstrap intervals.}
\label{tab:matched}
\centering\small
\setlength{\tabcolsep}{5pt}
\begin{tabular}{@{}lllrrrr@{}}
\toprule
Task & Path & Replacement & Replacement RMSE & Reduction & Pairs & RMSE gain interval \\
\midrule
Radius (fm) & MI & Tanh & $0.019905 \pm 0.000115$ & -0.21\% & 2/5 & [-0.000933, +0.001183] \\
Radius (fm) & AM & Tanh & $0.023060 \pm 0.001017$ & +9.98\% & 5/5 & [+0.000406, +0.004981] \\
Radius (fm) & MI & SiLU & $0.020206 \pm 0.000086$ & +1.28\% & 5/5 & [-0.000335, +0.001011] \\
Radius (fm) & AM & SiLU & $0.022337 \pm 0.001027$ & +7.07\% & 5/5 & [-0.000314, +0.004036] \\
\bottomrule
\end{tabular}
\end{table*}

Table~\ref{tab:matched} tests the radius PU mapping through the matched replacements in Section~\ref{sec:controls}. Both retain the MI branch; AM also retains its additive candidate and fusion gate. Candidate inputs, surrounding recurrence, shared-tensor initialization, and optimization remain fixed.

For radii, AM's product mapping reduces RMSE by 9.98\% relative to Tanh and 7.07\% relative to SiLU, improving all five pairs in both comparisons. MI-PU reduces mean RMSE relative to SiLU by 1.28\%, with lower error in all five pairs. Its RMSE is closely matched by MI-Tanh (Table~\ref{tab:radius}). Together, these results show how candidate fusion shapes the benefit of the learned product mapping.

Full/matched parameter counts are 10,085/10,022 real components for MI and 10,055/10,002 for AM; the differences are PU thresholds. Replacements remove the complete PU module, including its magnitude floor and logarithmic/exponential mapping.

\begin{table*}[t]
\caption{Paired radius-error gains (fm), comparator minus MI, with 95\% isotope-chain bootstrap intervals. Positive gains favor MI.}
\label{tab:paired}
\centering\small
\setlength{\tabcolsep}{5pt}
\begin{tabular}{@{}llrrrr@{}}
\toprule
Task & Candidate vs comparator & Reduction & $\Delta$RMSE & RMSE gain interval & MAE gain interval \\
\midrule
Radius (fm) & MI vs Real GRU & +2.21\% & $+0.000452$ & [-0.000642, +0.002006] & [-0.000860, +0.001691] \\
Radius (fm) & MI vs MLP & +6.41\% & $+0.001367$ & [-0.000701, +0.003376] & [-0.000576, +0.002651] \\
Radius (fm) & MI vs WS$^*$ prior & +14.40\% & $+0.003355$ & [+0.001569, +0.005167] & [+0.001219, +0.004468] \\
Radius (fm) & MI vs CV GRU & +3.82\% & $+0.000792$ & [-0.000530, +0.002690] & [-0.000570, +0.002101] \\
\bottomrule
\end{tabular}
\end{table*}

\subsection{Paired Radius Errors and Robustness of the Observed Gains}\label{sec:paired}

\begin{figure*}[t]
\centering
\textcolor{black}{\includegraphics[width=0.93\textwidth]{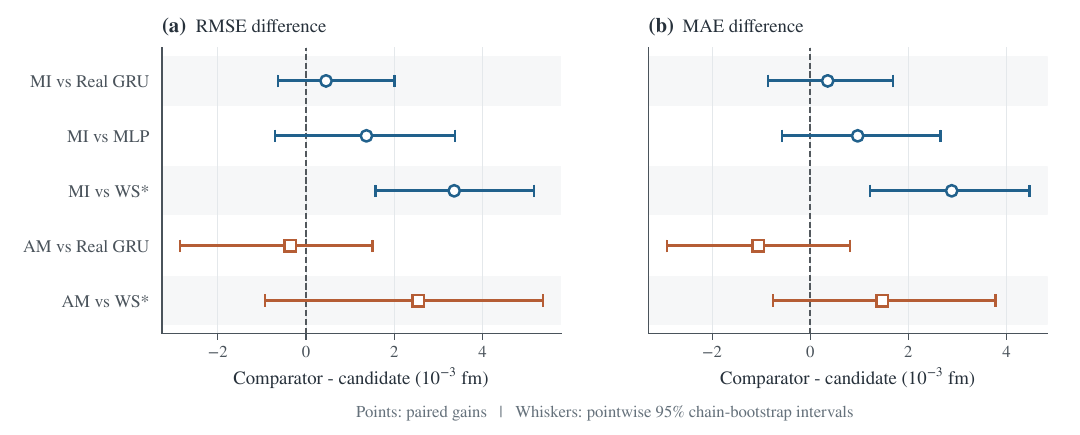}}
\caption{ Paired radius-error gains. MI and AM denote the full complex PU-GRU models. Points show comparator minus candidate errors; whiskers show pointwise 95\% isotope-chain bootstrap intervals. The dashed line marks zero gain.}
\label{fig:paired}
\end{figure*}

Table~\ref{tab:paired} reports paired radius gains for MI, while Fig.~\ref{fig:paired} shows selected MI and AM comparisons. Seed pairs reflect training variability; isotope-chain resampling assesses sensitivity to test-set composition.

MI reduces RMSE relative to WS$^*$ by 0.003355 fm, with a positive 95\% interval of $[0.001569,0.005167]$ fm. Intervals against both GRU baselines and MLP include zero. Among matched controls, the AM--Tanh RMSE gain interval is positive, whereas both SiLU intervals include zero (Table~\ref{tab:matched}).

\begin{figure*}[t]
\centering
\textcolor{black}{\includegraphics[width=0.98\textwidth]{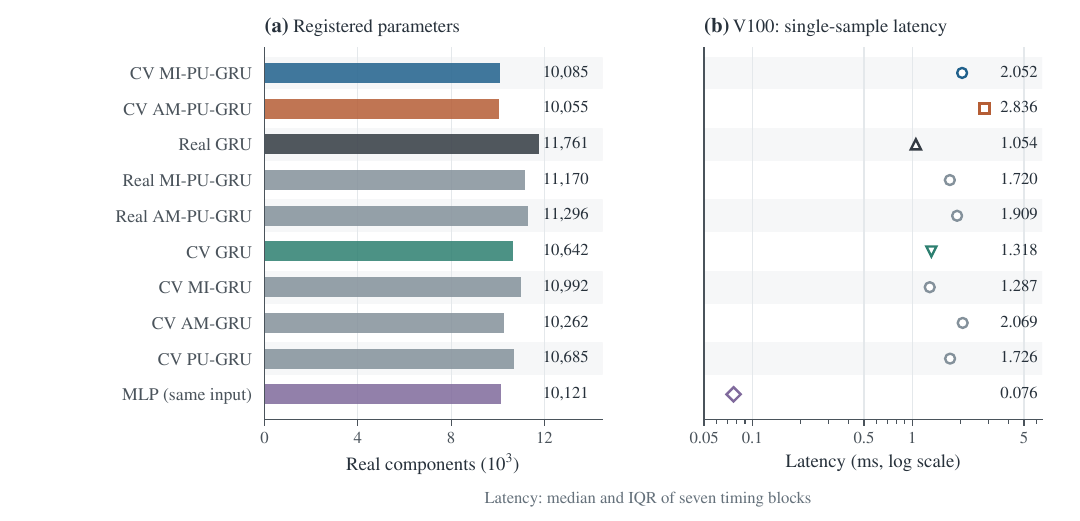}}
\caption{ Radius-model parameter counts and single-sample inference latency. Latency markers and whiskers show medians and interquartile ranges across seven block means; the latency axis is logarithmic.}
\label{fig:cost}
\end{figure*}

\subsection{Capacity and Inference Cost}\label{sec:cost}

Table~\ref{tab:cost} and Fig.~\ref{fig:cost} compare radius-model capacity and inference cost. MI and AM have about 10,000 real parameter components, comparable to MLP and fewer than real GRU. FLOPs count affine operations; latency also captures nonlinear and elementwise computation.

\begin{table*}[t]
\caption{ Radius-model capacity and inference cost on a Tesla V100-SXM2-32GB. Complex parameters count as two real components. FLOPs cover executed affine maps; latencies are medians of seven synchronized block means.}
\label{tab:cost}
\centering\small
\setlength{\tabcolsep}{5pt}
\begin{tabular}{@{}lrrrr@{}}
\toprule
Model & Real components & Linear FLOPs/sample & $B=1$ (ms) & $B=64$ (ms/batch) \\
\midrule
CV MI-PU-GRU & $10,085$ & $195,542$ & $2.0524$ & $2.0911$ \\
CV AM-PU-GRU & $10,055$ & $194,452$ & $2.8364$ & $2.8877$ \\
\midrule
Real GRU & $11,761$ & $115,321$ & $1.0540$ & $1.0812$ \\
Real MI-PU-GRU & $11,170$ & $107,416$ & $1.7200$ & $1.7347$ \\
Real AM-PU-GRU & $11,296$ & $108,947$ & $1.9092$ & $1.9781$ \\
\midrule
CV GRU & $10,642$ & $207,922$ & $1.3180$ & $1.3210$ \\
CV MI-GRU & $10,992$ & $160,582$ & $1.2873$ & $1.2993$ \\
CV AM-GRU & $10,262$ & $199,442$ & $2.0691$ & $2.0997$ \\
CV PU-GRU & $10,685$ & $207,922$ & $1.7260$ & $1.7518$ \\
\midrule
MLP (same input) & $10,121$ & $20,057$ & $0.0764$ & $0.0801$ \\
\bottomrule
\end{tabular}
\end{table*}

Inference is timed on a Tesla V100-SXM2-32GB with PyTorch 2.7.1 and CUDA 11.8. After 100 warm-up calls, seven blocks of 200 calls are timed with CUDA synchronization at batch sizes 1 and 64. Reported latencies are the medians of the seven block means. GPU-resident inputs are used in eager execution without mixed precision or CUDA Graph capture; timing covers the forward pass only. A real multiply--accumulate counts as two FLOPs and a complex one as eight; bias additions are counted separately.

Single-sample latency is 2.0524 ms for MI and 2.8364 ms for AM, versus 1.0540 ms for real GRU and 0.0764 ms for MLP. PU models trade additional nonlinear computation for predictive accuracy. Registered counts include 2,730 inactive optional components in MI-GRU; these are excluded from executed FLOPs.

\section{Discussion}\label{sec:discussion}

Beyond comparisons among our proposed variants, it is important to contextualize the performance of the PU-GRU architectures with respect to existing approaches in the literature. We summarize representative results reported in prior studies and compare them against our complex MI-PU-GRU and AM-PU-GRU models for masses and charge radii.

Partitions determine both evaluation targets and measured-neighbor inputs. Different partitions can therefore improve or worsen errors and change model rankings, including for shared targets whose neighbors differ.

The CV GRU baseline applies Tanh to both candidate components (Eq.~\eqref{eq:cvgru}), whereas the conference implementation~\cite{li2026conference} used $\tilde h_t=\tanh(\Re b_t)+\mathrm{i}0$. MI-PU-GRU and AM-PU-GRU retain their original architectures.

\subsection{Mass prediction in the literature}\label{sec:literature_known}

Table~\ref{tab:lit_t1} groups literature and present-study mass errors by evaluation protocol and lists their nuclear selections and inputs.

\begin{table*}[t]
\centering
\caption{ Literature and present-study mass errors (MeV) for known nuclei, grouped by evaluation protocol.}
\label{tab:lit_t1}
\small
\setlength{\tabcolsep}{5pt}
\begin{tabularx}{\textwidth}{@{}>{\raggedright\arraybackslash}p{3cm}>{\raggedright\arraybackslash}X >{\raggedright\arraybackslash}p{5.5cm}r@{}}
\toprule
Method & Evaluation scope & Representation / setting & RMSE \\
\midrule
\multicolumn{4}{@{}l}{\textit{Published held-out evaluations}}\\
MISR~\cite{munoz2025discovering} & $12\le Z\le50$; 20\% test split & Ten-term symbolic expansion & 0.99 \\
CatBoost~\cite{guo2025probing} & UNEDF1-HFB coverage; 80/20 split & HFB binding-energy correction & 0.189 \\
PI-FCNN~\cite{huang2025validation} & Case A; $Z,N>20$, uncertainty $<100$ keV & Physics-informed Method III & 0.122 \\
CPUN-L~\cite{dellen2024masses} & Test-A: experimental masses; $A\ge20$, uncertainty $\le0.3$ MeV & Complex feedforward PU; five-trial mean & 0.438 \\
RNN~\cite{jalili2025recurrent} & AME2020, $Z,N\ge8$; 90/5/5 split & Eleven features, Tanh & 0.6016 \\
LSTM~\cite{jalili2025recurrent} & Same published protocol as RNN & Eleven features, Tanh & 0.5574 \\
GRU~\cite{jalili2025recurrent} & Same published protocol as RNN & Eleven features, Tanh & 0.4591 \\
\midrule
\multicolumn{4}{@{}l}{\textit{Whole-coverage evaluations, including training nuclei}}\\
MDN~\cite{mumpower2023bayesian} & AME2020 evaluation coverage, $A\ge50$ & Mixture-density mass model & 0.395 \\
MDN~\cite{li2024atomic} & AME2020 evaluation coverage, $Z\ge10$ & HFB32-augmented training & 0.246 \\
\midrule
\multicolumn{4}{@{}l}{\textit{Binding-energy regressions: regional or summary metrics}}\\
XGBoost~\cite{pandey2024prediction} & AME2020; $50<A<150$ (Table II) & Binding-energy regression & 3.702 \\
MLP regressor~\cite{pandey2024prediction} & AME2020; $A<50$ (Table II) & Binding-energy regression & 3.128 \\
Random forest~\cite{pandey2024prediction} & AME2020; overall value reported in conclusion & Binding-energy regression & 3.089 \\
\midrule
\multicolumn{4}{@{}l}{\textit{Present study: Mass I}}\\
CV MI-PU-GRU & $Z,N\ge8$; 70/15/15 split & Measured-neighbor sequence + WS4 & $0.2708$ \\
CV AM-PU-GRU & Same partitions and inputs & Measured-neighbor sequence + WS4 & $0.2698$ \\
\bottomrule
\end{tabularx}
\par\smallskip
\begin{minipage}{\textwidth}\footnotesize
The published RNN/LSTM/GRU evaluation includes AME-extrapolated entries, which our selection excludes. Binding-energy rows use regional or conclusion-level metrics.
\end{minipage}
\end{table*}

As shown in Table~\ref{tab:lit_t1}, Jalili et al.~\cite{jalili2025recurrent} reported results for recurrent architectures, including GRU, RNN, and LSTM, but their models utilize broader input features, with eleven features per nucleus, and standard recurrent designs without product-unit augmentation. Our complex-valued PU-GRU models use a physics-informed neighborhood representation: three channels at five positions, comprising $Z$, $N$, and a measured neighbor mass or target WS4 estimate.

We also compare with the CPUN-L proposed by Dellen et al.~\cite{dellen2024masses}, which yields an RMSE of 0.438 MeV. While their model also incorporates product units and complex-valued representations, it is based on a fully connected feedforward architecture. In contrast, our PU-GRU framework integrates multiplicative interactions into a recurrent paradigm and explicitly exploits sequence-level dependencies among neighboring nuclides.

Categorical gradient boosting (CatBoost)~\cite{guo2025probing} and physics-informed fully connected neural networks (PI-FCNN)~\cite{huang2025validation} report lower errors of 0.189 and 0.122 MeV. CatBoost corrects UNEDF1-HFB predictions; PI-FCNN uses a physics-informed correction for nuclei with $Z,N>20$ and mass uncertainties below 100 keV.

\subsection{Temporal and release-transfer comparisons} \label{sec:literature_temporal}
\begin{table*}[t]
\centering
\caption{ Reference-set and later-set mass RMSEs (MeV), grouped by evaluation protocol. Dashes indicate unreported values.}
\label{tab:lit_t2}
\small
\setlength{\tabcolsep}{3pt}
\begin{tabularx}{\textwidth}{@{}>{\raggedright\arraybackslash}p{3.5cm}>{\raggedright\arraybackslash}X rr@{}}
\toprule
Method & Evaluation and input context & Reference RMSE & Later-set RMSE \\
\midrule
\multicolumn{4}{@{}l}{\textit{Held-out reference / later-measurement evaluations}}\\
GPR~\cite{yuksel2024nuclear} & Twelve features; $Z,N\ge8$; reference hold-out and newly measured AME2020 nuclei & 0.26 & 0.67 \\
SVR~\cite{yuksel2024nuclear} & Same published protocol as GPR & 0.39 & 0.74 \\
CNN--WS4~\cite{lu2025nuclear} & $5\times5$ nuclear grid; WS4 residual correction; new AME2020 measurements & --- & 0.211 \\
PI-FCNN~\cite{huang2025validation} & Case B, Method III; $Z,N>20$ and uncertainty $<100$ keV; AME2016 to AME2020 & --- & 0.191 \\
\midrule
\multicolumn{4}{@{}l}{\textit{Whole evaluation coverage, including training nuclei}}\\
MDN~\cite{mumpower2022masses} & $Z\ge20$; entire evaluated AME2016 and AME2020 coverage & 0.316 & 0.336 \\
\midrule
\multicolumn{4}{@{}l}{\textit{Present study: Mass II}}\\
CV MI-PU-GRU & Interpolation and temporal test sets & $0.2536$ & $0.2251$ \\
CV AM-PU-GRU & Same partitions and inputs as MI & $0.2509$ & $0.1905$ \\
\bottomrule
\end{tabularx}
\par\smallskip
\begin{minipage}{\textwidth}\footnotesize
GPR/SVR use AME2016 membership to define the reference pool. PU-GRU uses discovery-year groups and a fixed WS4 prior; both use AME2020 mass labels.
\end{minipage}
\end{table*}

Table~\ref{tab:lit_t2} distinguishes later-measurement tests from whole-evaluation errors. GPR/SVR, CNN--WS4, and PI-FCNN use their respective later-measurement selections; MDN reports errors over the full evaluated AME2016 and AME2020 coverage~\cite{mumpower2022masses}.

PU-GRU uses the same neighborhood construction for reference-pool and temporal targets. AM's lowest mean interpolation and temporal errors among the configurations in Table~\ref{tab:mass} support combining an additive candidate with the learned product representation.

\subsection{Charge-radius prediction in the literature}\label{sec:literature_radius}

\begin{table*}[t]

\centering
\caption{ Charge-radius RMS errors (fm), grouped by evaluation protocol. Ranges span published model variants; present neural-model results are five-seed means $\pm$ sample standard deviations; the prior is deterministic.}
\label{tab:lit_radius}
\small
\setlength{\tabcolsep}{5pt}
\begin{tabularx}{\textwidth}{@{}>{\raggedright\arraybackslash}p{2.5cm}>{\raggedright\arraybackslash}X >{\raggedright\arraybackslash}p{6cm}r@{}}
\toprule
Method & Evaluation scope & Physical input / learned representation & RMSE (fm) \\
\midrule
\multicolumn{4}{@{}l}{\textit{Published later-measurement evaluations}}\\
DFT--BNN~\cite{utama2016radius} & $Z\ge20$, $A\ge40$; learning set and later measurements & Residual refinement of NL3, FSUGold, and FSUGarnet; $(Z,A)$ inputs & 0.0280--0.0327 \\
NP--BNN (D6)~\cite{dong2023radius} & $Z\ge20$, $A\ge40$; training set and later measurements & Formula residual; six inputs including pairing, shell, isospin, and Hg-shape features & 0.0139 \\
Formula--KRR~\cite{ma2022radius} & $Z,N\ge8$; CR2004 training; later CR2013 measurements & Six phenomenological formulas with Gaussian-kernel residual correction & 0.0268--0.0369 \\
\midrule
\multicolumn{4}{@{}l}{\textit{Published leave-one-out evaluations used for hyperparameter selection}}\\
Formula--KRR~\cite{ma2022radius} & $Z,N\ge8$ & Same six formula families; refitted physical parameters & 0.0165--0.0197 \\
RCHB--EKRR~\cite{ma2024radius} & $Z\ge8$ & RCHB residual; Gaussian and odd--even kernels & 0.0092 \\
WS$^*$--EKRR~\cite{ma2024radius} & Same leave-one-out protocol & Five-parameter WS$^*$ radius formula; odd--even residual kernel & 0.0096 \\
\midrule
\multicolumn{4}{@{}l}{\textit{Published global calibration}}\\
WS$^*$ radius formula~\cite{wsradius} & Measured radii; $A\ge16$ & Four-parameter formula with shell corrections and deformations & 0.022 \\
\midrule
\multicolumn{4}{@{}l}{\textit{Present study: held-out network evaluation under a fixed prior}}\\
CV MI-PU-GRU & $Z,N\ge8$; 70/15/15 & Four measured training neighbors + fixed four-parameter WS$^*$ prior & $0.019947\pm0.000121$ \\
CV AM-PU-GRU & Same partition and inputs & Additive--multiplicative candidate fusion & $0.020758\pm0.000517$ \\
WS$^*$ prior alone & Same test nuclei & Fixed prior, without neural prediction & 0.023303 \\
\bottomrule
\end{tabularx}
\par\smallskip
\begin{minipage}{\textwidth}\footnotesize
CR2004 and CR2013 denote the radius compilations used in Formula--KRR. Its physical parameters were fitted to the full evaluation pool, and later-measurement predictions reused leave-one-out hyperparameters. The WS$^*$ rows use different formula versions and nuclear selections.
\end{minipage}
\end{table*}

Table~\ref{tab:lit_radius} compares physical and learned radius predictors. Bayesian networks refine density-functional models~\cite{utama2016radius} or phenomenological formulas using pairing, shell, isospin, and shape features~\cite{dong2023radius}. Feedforward networks also incorporate quadrupole-transition and symmetry-energy information~\cite{wu2020radius}.

Kernel methods model correlations among nuclei through phenomenological-formula corrections~\cite{ma2022radius} and odd--even kernels~\cite{ma2024radius}. The latter reach leave-one-out errors of 0.0092 fm for RCHB--EKRR and 0.0096 fm for WS$^*$--EKRR. Their WS$^*$ baseline has five parameters; ours uses the original four-parameter formula~\cite{wsradius}.

PU-GRU predicts radii directly from measured neighborhoods and a theoretical estimate, rather than fitting formula residuals. Comparisons with GRU and MLP evaluate this recurrent representation under identical inputs.

\subsection{Product-unit modeling across nuclear observables}

Mass comparisons favor the full PU-GRU designs across the configurations and candidate replacements in Tables~\ref{tab:mass} and~\ref{tab:mass_checkpoint} under the WS4 prior, although MI-PU-GRU and MI-SiLU have similar mean temporal errors. Radius controls show that the benefit of the product mapping depends on candidate construction.

AM combines additive and product pathways, and both matched radius replacements increase its mean error. MI supplies explicit interaction features to the PU and has lower mean radius RMSE than both GRU baselines, while its Tanh replacement achieves a slightly lower RMSE. These results favor learned products in suitable candidate pathways, with the benefit varying between MI and AM.

Physical priors provide global nuclear systematics, while recurrence integrates local measurements. Error-distribution and inference-cost analyses show where this representation improves predictions and what computation it requires.

\section{Conclusion}\label{sec:conclusion}

We have presented complex-valued extensions of the MI-PU-GRU and AM-PU-GRU architectures for nuclear mass and charge-radius prediction. By embedding multiplicative interactions and product-unit transformations into recurrent units and extending them to the complex domain, our models capture nonlinear dependencies among neighboring nuclei while retaining gated control of the recurrent state.

Through systematic evaluation on interpolation and extrapolation tasks, we have shown that the complex AM-PU-GRU model achieves the lowest mean mass RMSE among the models in Tables~\ref{tab:mass} and~\ref{tab:mass_checkpoint} under the WS4 prior, across the three data partitions and three training seeds per partition. Comparative analyses further reveal that the complex PU-GRU variants have lower mean mass errors than their real-valued counterparts, while configuration and candidate-replacement comparisons support the complete MI--PU and AM--PU designs.

For charge radii, MI reduces the WS$^*$ prior RMSE by 14.40\% without mass features. AM's matched controls support combining product and additive pathways, extending the framework to a second nuclear observable.

Future work will explore extending complex PU-based recurrent networks to other nuclear observables and to broader domains where extrapolation capability is essential.

\section*{CRediT authorship contribution statement}
\textbf{Ziyuan Li:} Conceptualization, Methodology, Software, Data curation, Investigation, Formal analysis, Validation, Visualization, Writing -- original draft, Writing -- review \& editing.
\textbf{Paulo S. A. Freitas:} Conceptualization, Methodology, Formal analysis, Validation, Writing -- original draft, Writing -- review \& editing.
\textbf{John W. Clark:} Conceptualization, Methodology, Formal analysis, Validation, Writing -- original draft, Writing -- review \& editing.
\textbf{Babette Dellen:} Conceptualization, Methodology, Formal analysis, Validation, Writing -- original draft, Writing -- review \& editing.

\section*{Declaration of competing interest}
The authors declare no competing financial interests or personal relationships that could have influenced the work reported in this paper.

\section*{Acknowledgements}
We thank Amir Jalili (Zhejiang Sci-Tech University) for providing the nuclear-property compilation used in this study.

\section*{Data availability}
The third-party compilation is not publicly redistributed. Original sources identify the evaluated masses, experimental radii, and theoretical priors~\cite{ame2020i,ame2020ii,angeli2013,ws4,wsradius}. Selection, preprocessing, neighborhood construction, and experimental settings are specified in the article. The project repository is
\url{https://github.com/ziyuan-li/pu_gru}.

\section*{Declaration of generative AI use}
During the preparation of this manuscript, the authors used OpenAI ChatGPT to assist with language editing, manuscript organization, and literature retrieval. After using this tool, the authors reviewed and edited the content as needed and take full responsibility for the content of the published article.

\bibliographystyle{cas-model2-names}
\bibliography{references}
\end{document}